# Enhancement of graphene visibility on transparent substrates by refractive index optimization

Hugo Gonçalves,[1] Luís Alves,[1] Cacilda Moura,[1] Michael Belsley,[1] Tobias Stauber,[2] and Peter Schellenberg[1,*]

[1]*Department of Physics, University of Minho, Campus de Gualtar, Pt-4710-057 Braga, Portugal*
[2]*Department of Condensed Matter Physics, University Autónoma de Madrid, Campus de Cantoblanco, E-28049 Madrid, Spain*
[*]*peter.schellenberg@fisica.uminho.pt*

**Abstract:** Optical reflection microscopy is one of the main imaging tools to visualize graphene microstructures. Here is reported a novel method that employs refractive index optimization in an optical reflection microscope, which greatly improves the visibility of graphene flakes. To this end, an immersion liquid with a refractive index that is close to that of the glass support is used in-between the microscope lens and the support improving the contrast and resolution of the sample image. Results show that the contrast of single and few layer graphene crystals and structures can be enhanced by a factor of 4 compared to values commonly achieved with transparent substrates using optical reflection microscopy lacking refractive index optimization.

## 1. Introduction

Graphene is a novel material that has been attracting widespread interest due to its unique electronic, optical, magnetic, and mechanical properties [1-4]. Graphene's outstanding characteristics make it extremely appealing for a wide range of applications. In electronics, graphene, which has a zero band-gap, has been used to create transistors [2], while its versatility has been increased using several different approaches to engineer a band gap in this material [5,6]. Graphene is also used in spintronics [7], in new hybrid materials for biomedical systems [8], to produce gas and bio sensors [9,10], electrodes [11,12], transparent electrodes for solar cells and LCD displays [13], supercapacitors [14], and as a nonlinear element in laser applications [15].

For many of these investigations robust and easily applicable imaging methods with resolution in the micrometer range is obligatory. Optical reflection microscopy is a simple, high-throughput technique that can be used to determine whether single layers are present, to measure their sizes and positions, and to determine the quality of the samples. Due to the low reflectivity of single layer graphene, interference techniques utilizing dielectric-coated wafers as a substrate to obtain high-contrast images using optical reflection microscopy were first introduced by Blake et. al. [16] and later modified [17-21]. Ellipsometry [22,23], phase-shifting interferometric imaging [24], surface plasmon resonance reflectance [25] and Rayleigh [26] and Raman imaging microscopy [27-29] have also been applied to identify graphene layers deposited on different types of substrates. Further methods exploit surface hydrophobicity [30] or quenching of dye molecules by graphene crystals [31].

The optical observation of graphene layers on transparent substrates would be an asset due to the versatility and variability of these materials. However, the contrast is typically quite small due to the low optical conductivity value of a single graphene sheet [32]. For example even though only 4% of the incident light is typically reflected from a glass substrate in the visible, the optical conductivity of a single layer of graphene is such that a contrast of only 7% is obtained in the visible [33], which makes the observation of graphene on transparent substrates notoriously difficult. To overcome this obstacle, a novel technique is presented here to enhance the visibility of graphene monolayers using optical reflection microscopy; by introducing a medium with a refractive index tuned sufficiently close to that of the substrate, the optical contrast of graphene flakes can be greatly enhanced. Using this method we have obtained graphene image contrasts that approach 30%, roughly 4 times higher than values typically reported for a graphene monolayer on a glass support, and 2 times higher than the contrast observed in interference techniques. Even higher contrast values are achievable in principle by further optimization of the refractive index, accompanied by a lowering of the intrinsic noise of the detection system.

## 2. Theoretical basis

The experiments are based on optical reflection microscopy in combination with an immersion medium of the same refractive index as the substrate between the sample and the front lens of the objective. In the ideal case, there will be no reflection from the substrate surface, and only the reflection from single -and multilayer graphene will be visible. The Contrast $C$, which establishes the relative difference between the reflected light intensity with $m$ graphene-layers $I(m)$ and with no graphene layer $I(m=0)$, is described by:

$$C = \frac{I(m) - I(m=0)}{I(m=0)} \qquad (1)$$

The reflectivity with and without graphene is given by the Fresnel coefficients for linearly polarized light which for normal incidence are given by [32]:

$$r = \frac{n_1 - n_2 - m\pi\alpha}{n_1 + n_2 + m\pi\alpha} \qquad (2)$$

where $n_1$ and $n_2$ denote the refractive index of the medium above and below the graphene sample, $m$ the number of graphene layers and $\alpha=1/137$ the fine-structure constant. The intensity is then given by $I = |r|^2$. Equation (1) thus gives the following expression:

$$C = \frac{4mn_1\pi\alpha(-n_1^2 + n_2(n_2 + m\pi\alpha))}{(n_1 - n_2)^2(n_1 + n_2 + m\pi\alpha)^2} \qquad (3)$$

This expression formally diverges for $n_1=n_2$ which is an artifact of our simplistic theory since there will always be some light background from the substrate, for example due to light scattering in the liquid, due to noise of the camera system, or averaging over a small range of incident angles due to the finite numerical aperture of the objective lens. Nevertheless, the above formula exemplifies the effect which we intent to exploit, i.e., choosing an immersion liquid with refractive index $n_1$ approximately equal to that of the substrate $n_2$ can significantly increase the sample contrast. The above formula also states that for $n_1 \neq n_2$ the contrast approximately scales with the number of layers since $m\alpha=m/137$ cannot be neglected in comparison with $n_1$ and $n_2$. This will also experimentally be confirmed.

To account for the various sources of residual reflections mentioned above we have chosen to introduce a phenomenological reflection constant $R$, which essentially indicates the level of the background signal that would be observed from the substrate under conditions of perfect index matching. This constant term should be independent of the presence of graphene and thus cancels in the numerator of Eq. (1). But in the denominator, it will lead to finite results. Our amended contrast formula thus reads:

$$C = \frac{4mn_1\pi\alpha(-n_1^2 + n_2(n_2 + m\pi\alpha))}{((n_1 - n_2)^2 + (n_1 + n_2)^2 R)((n_1 + n_2 + m\pi\alpha)^2)} \quad (4)$$

The theoretical predictions for the contrast for glycerol, oil and quinoline ($n_1$=1.47, $n_1$=1.49 and $n_1$=1.63, respectively) on glass ($n_2$=1.52) compare well with the experimental values if we choose $R$=0.0007. For single-layer graphene, we have $C$ ($n_1$=1.47) ≈31 %, $C$ ($n_1$=1.49) ≈26 % and ($n_1$=1.63) ≈-26 %, for double-layer graphene, we find $C$ ($n$=1.47) ≈73 %, $C$ ($n_1$=1.49) ≈65 % and $C$ ($n_1$=1.63) ≈-43 %. The value of $R$ will change slightly with experimental conditions, but it will be always of this magnitude and can be neglected if $I(0)$>$R$, i.e., if the refractive index of the two media considerably differ from one another, which is, for example, the case of an air-glass interface. Figure 1 presents the results applying Eq. (4) which includes $R$. Figure 1(a) shows the change of contrast as a function of the refractive index of the medium between the microscope objective lens and graphene layers. In particular the contrast reaches its maximum at a refractive index value smaller than $n_2$, which would have not been predicted by using Eq. (3) without assuming residual reflection. Furthermore, a negative contrast, which would be visible as a darker graphene layer in front of a brighter substrate background, is predicted for refractive indexes above $n$=1.53 and shows a maximum in the negative around $n$=1.63, which will be verified experimentally. This analysis is also consistent with similar observations using mica as a dielectric on top of a graphene sheet [34].

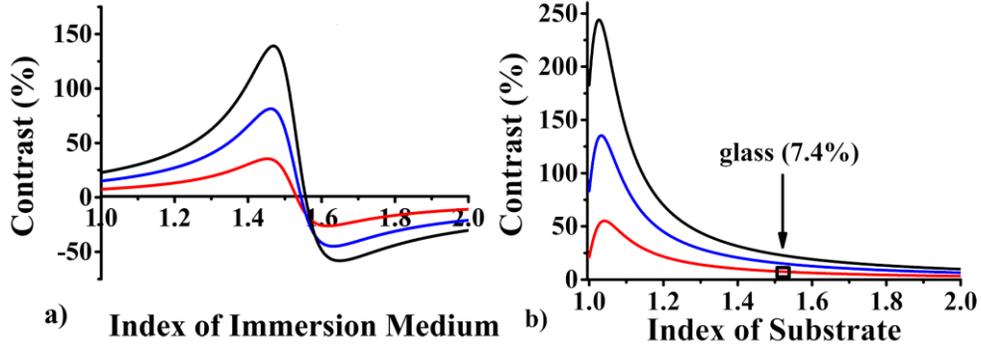

Fig. 1(a) The contrast as a function of the immersion index medium for a monolayer (red), bilayer (blue) and trilayer (black) deposited on a glass substrate. Fig. 1(b) The contrast of graphene layers as a function of the substrate index in air.

Figure 1(b) shows the contrast as a function of the refractive index of the support according to Eq. (4), and for values sufficiently different from $n$=1, is in line with data published before [33]. Specifically, the contrast rises significantly upon lowering the refractive index, however there are no solid transparent materials available with a refractive index close to one. To conclude, the theory can predict qualitatively as well as quantitatively the optimum conditions for the optical contrast for a single layer or few layers of graphene in a given system with a fixed index of refraction of the support

## 3. Experimental details

Graphene samples were prepared by micromechanical cleavage (also known as the scotch-tape technique) [35] of 5-10 mm graphite flakes from NGS Naturgraphit and subsequently transferred to standard glass slides ($n$=1.52). The samples were observed under a Nikon

Optiphot metallurgical reflection microscope in 20× magnification. The objective is specified for use with an immersion liquid with a refractive index around n=1.5. Figure 2 illustrates the experimental set-up that shows how refractive index optimization is used to enhance the contrast of graphene flakes. Additionally a prism was firmly attached to the bottom surface of the glass slide using immersion oil to minimize any reflections from the second refractive index boundary.

The refractive index solutions were filtered through a Whatman Puradisc membrane syringe filter with 450 nm pore size to remove any suspended impurity particles whose presence would lead to a larger scattering from the liquid medium thereby increasing the background.

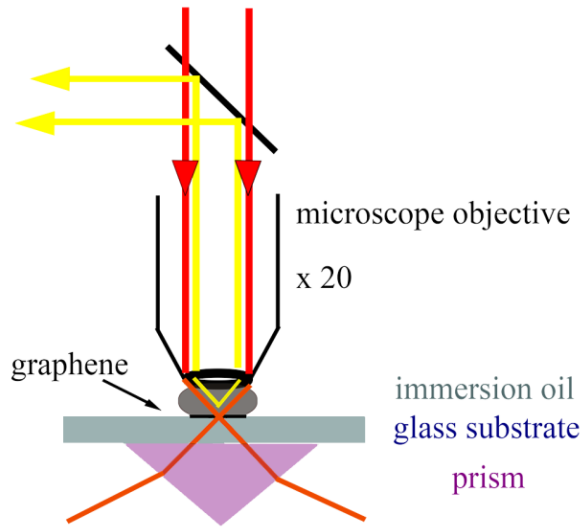

Fig. 2 Schematics of the optical reflection microscope set-up with the addition of immersion oil between the glass and the graphene and a prism to prevent reflections from the back of the substrate. For clarity the reflected beam is drawn in a different colour (yellow) compared to the incoming beam (orange).

Figure 3 shows images of the same graphitic layers obtained in air (a), glycerol (b), immersion oil (c), and quinoline (d). The brightest region in the photograph is bulk graphite, the transparent regions are graphene monolayers and bilayers as established by Raman microscopy. The use of glycerol ($n=1.47$) and oil ($n=1.49$) enhances the contrast significantly. The flakes appear bright on a darker background, which is the glass support in the absence of any graphene. The contrast of the graphene layers in quinoline ($n=1.63$) as a medium is also enhanced, however the flakes appear darker compared to the background, in line with the theoretical predictions of Eq. (4).

For quantification, Figure 4 shows the contrast profile as a function of the position along the blue line shown in the image of Fig. 3(c) for all the immersion materials. The contrast values (in percent) are plotted for each pixel against the averaged light intensity from the glass support, according to Eq. (1). The images obtained using glycerol and oil show a contrast that is approximately 4 times larger than the contrast for the image acquired with the sample exposed to air, while the contrast using quinoline is increased by a factor of ~3, but with a reversed sign. This confirms the theoretical predictions presented above.

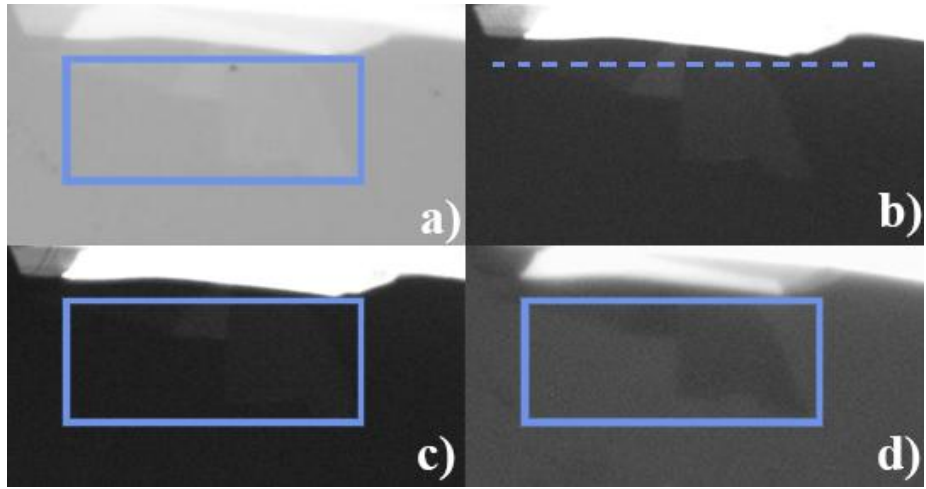

Fig. 3 Images of the same graphene flake in air (a), glycerol (b), immersion oil (c), and quinoline (d). The broken blue line shown in image b indicates the positions for the profile taken in each of the images to create Fig. 4. The blue rectangle represents the region used in Fig. 5 for the contour maps.

To the best of our knowledge, the typical contrast value previously reported for graphene monolayers [17,33,36] obtained with glass substrates is 7%, compared to the 30% reported here for a monolayer of graphene. In order to verify the correspondence between the contrast and the number of graphene layers in the different regions of the flake, Raman microscopy was employed for confirmation [29,37,38]. The spectra were taken in three different locations of the transparent regions of the flakes inside the square marked in each image shown in Fig. 3. From the analyses of the G and D peaks in the Raman spectra (not shown) it can be concluded that the regions with the lowest reflectance in the images are graphene monolayers, and the area with twice the contrast in-between these 2 regions is a graphene bilayer.

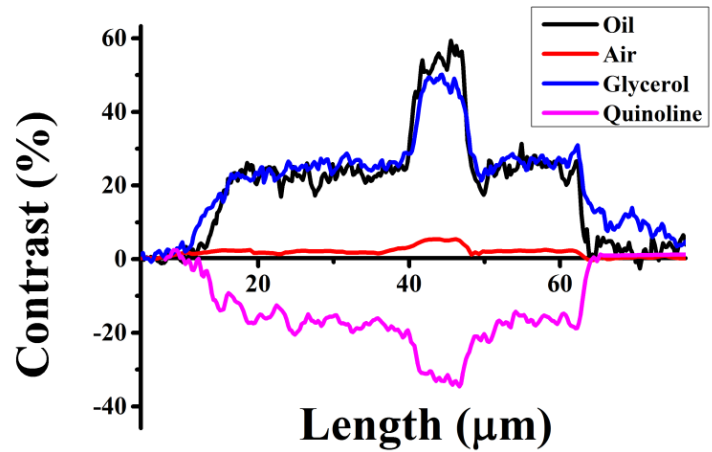

Fig. 4 The graphs displays a comparison of the contrast profile in the flake within the squares of the images presented in Fig. 3 for air, microscopy immersion oil, glycerol and quinoline.

In order to do perform a statistic analysis of the images of Fig. 3, several pictures were taken under the same conditions and the contrast was then averaged for each pixel in order to create the three-dimensional charts presented in Fig. 5. The contour map of the contrast using oil (b) presents a significantly better contrast and signal to noise ratio compared to air (a). The contour map from the analogue experiment using glycerol (not shown) is comparable to the one using oil. The contour map of quinoline (c) is qualitatively different, as the contrast is also enhanced relative to air, but is negative.

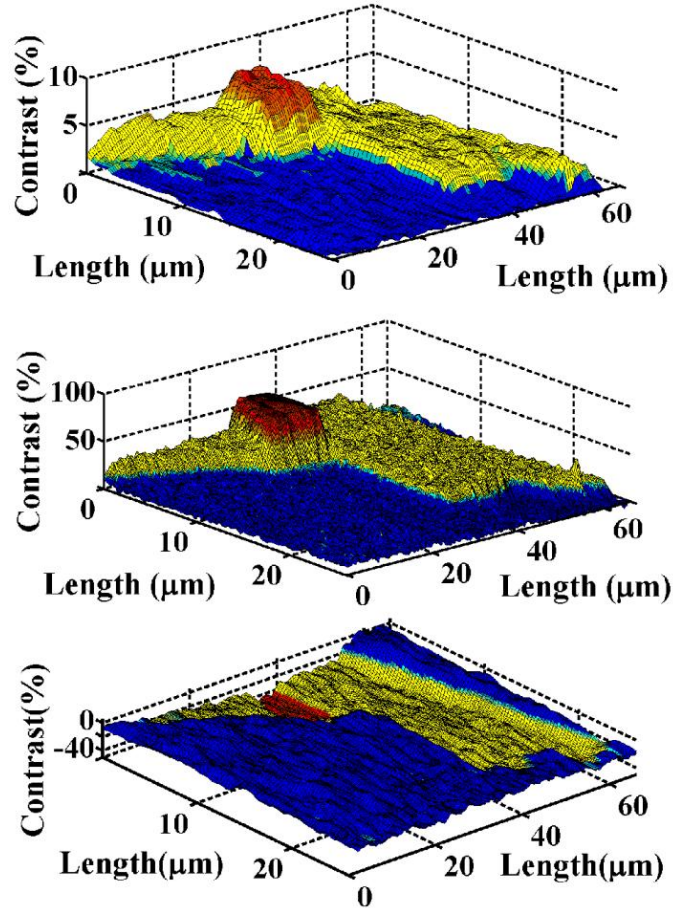

Fig. 5 Contour maps showing averaged contrast as a function of pixel position for the same region using air (a), oil (b) and quinoline (c) as a medium. The sample area depicted in these maps is indicated as squares in the images in Fig. 3. The bluish coloured areas represent the substrate baseline, while the yellow areas shows a monolayer and the red areas show a bilayer. The perspective of (a) and (b) is identical, the one of (c) is slightly altered to give an improved view of the negative contrast valley of the double layer graphene.

The contrast enhancement by this method is limited by light scattering in the immersion liquid, the noise of the camera system, and the reflection of light at the interface between the substrate and immersion liquid due to incomplete refractive index matching across the optical spectrum, all of which give a contribution to the phenomenological factor $R$ in Eq. (4). In principle, any given equipment could be standardized, as R could be determined independently of the presence of graphene by measuring the image illumination intensity for a set of immersion liquids of different, accurately determined refractive index values. In practice, it may be easier to simply measure the contrast of a graphene flake that is known to be a single layer and use Eq. (4) to estimate the value of R for a given experimental set-up. Of course the smaller the value of R is, the greater the improvement in contrast that can be achieved. Several measures can be taken towards this purpose. The immersion liquids should contain low levels of impurities, in particular they should be free of larger scattering particles and fluorophors. A smaller numerical aperture will help to reduce the background signal by limiting the range of incident illumination angles, since the precise index matching condition varies with incident angles. However this will inadvertently lead to a smaller spatial resolution. Similarly, reducing the spectral width of the illumination will lead to a smaller spread of the refractive indexes, but the price for this would be a weaker image illumination intensity. We have experimented with using different CCD-Cameras and different objectives, and have found that the effective R values is relatively robust to changes of the set-up.

## 4. Conclusions

Graphene monolayers deposited on glass substrates are notoriously difficult to see in optical reflection microscopes, as both the areas covered with graphene as well as the regions with exposed substrate reflect light and the consequently the contrast is low. Using a liquid which has a refractive index close to that of the substrate eliminates most of the reflection from this surface and as a consequence the visibility of graphene structures is greatly enhanced. Fresnel theory was used to compute the contrast as a function of the substrate and immersion medium indices of refraction and taking in account the optical conductivity of graphene. Experimental results show that by placing an immersion liquid in the space between the microscope lens and the glass substrate, the optical contrast can be improved by up to a factor of 4 relative to the ones obtained with the substrate exposed to air. In principal, even higher contrast could be achieved by reducing R, the level of residual light background from the substrate.

    The contrast of 30% for a monolayer of graphene on transparent substrates is twice as high compared to the standard microscopy technique exploiting interference enhancement [16-21] for which contrasts of 15% are observed. Furthermore, interference techniques require specific substrates, such as Si wafers coated with a dielectric of well defined thickness, and the contrast enhancement is strongly dependent on the wavelength of illumination. On the other hand, refractive index tuning can be employed with a large variety of transparent substrates, and there is hardly any variation of contrast with wavelength.

    Although we limited our investigation to graphene flakes from exfoliated graphite, this new method can be also used to visualize other graphene structures and patterns on transparent surfaces and therefore has potential to be applied on graphene-based devices.

**Acknowledgment**

We thank Ana Nicolau for experimental support and Nuno Peres and César Bernardo for useful discussions. This work was supported by FEDER through the COMPETE Program and by the Portuguese Foundation for Science and Technology (FCT) in the framework of the Strategic Project PEST-C/FIS/UI607/2011 and under grant PTDC/FIS/101434/2008. Luís Alves acknowledges support for an individual FCT postdoctoral grant under contract number SFRH/BPD/79842/2011. Tobias Stauber acknowledges support by Spain's Ministerio de Ciencias e Innovación (MICINN) via grant no. FIS2010-21883-C02-02.